\documentclass[11pt, oneside]{article}   	% use "amsart" instead of "article" for AMSLaTeX format
\usepackage{geometry}                		% See geometry.pdf to learn the layout options. There are lots.
\usepackage{graphicx}				% Use pdf, png, jpg, or eps§ with pdflatex; use eps in DVI mode
\usepackage{amssymb}
\usepackage{amsmath}
\usepackage[utf8]{inputenc}
\usepackage{multirow}
\usepackage{array}
\usepackage{mathtools}
\usepackage{makecell}
\usepackage{diagbox}
\usepackage{soul}
\usepackage[all,knot,poly]{xy}
\usepackage{color}
\numberwithin{equation}{section}

\begin{document}
\title{My Century of Physics}
\author{Robert J. Finkelstein}
\date{}
\maketitle
Bohr's first paper on the quantum theory of the hydrogen atom was published in 1913 and the General Theory of Relativity was published in 1916. I've had the good fortune of living through this historic last century of physics and, since I have been asked to say something about myself for this centennial party, I will sketch the intersection of certain points of my worldline with this big world of physics.
\newline

I became an undergraduate at Dartmouth College in 1933. The freshman course in physics was given by G.F. Hull who with E.L. Nichols, the founder of the Physical Review, was the first, with P. Lebedev in Russia to measure the pressure of light. (The world would learn sometime later about the diabolical importance of the pressure of light in enabling the hydrogen bomb.) Hull was a Canadian who had studied physics in Germany and had deep affection for his hosts, and so he began my first physics course with an emotional denunciation of what was happening to Germany in 1933.
\newline

After graduation I took my introductory course in Quantum Mechanics from I.I. Rabi at the Columbia summer school. My next course in this subject was given at Harvard by E.C. Kemble who had just finished one of the earliest books on Quantum Mechanics. I chose J.H. van Vleck as my thesis advisor. The first project that he offered me turned out to be significant enough to him to reference in his Nobel Prize Lecture (1979) for his pioneering and fundamental work in quantum magnetism. (The project that he had assigned to me in 1940 was the calculation of the energy levels of the chromium ion in a crystalline spectroscopic environment similar to that used by Maiman in his discovery of the optical laser in 1960.)
\newline

Before I received my doctoral degree Nazi submarines were dropping magnetic mines on the east coast to disrupt the seaborne oil traffic. I had by this time registered for the draft and van Vleck had sent me to Francis Bitter at MIT, who was assembling a group to devise countermeasures to the magnetic mines. So the day I took my last doctoral exam I also took the train to Washington to join Bitter's group in the Navy Dept. The magnetic mine problem was quickly solved and Bitter next began to assemble an operational research group with some leading mathematicians including Marshal Stone and Joseph Doob. I made one useful contribution during this period (to our submarine fleet), but felt I could do more elsewhere and succeeded in getting transferred to the group that dealt with shockwaves and detonation theory. There I was proud to have found an analytic solution to a shockwave problem that Chandrasekhar had previously solved numerically and I also wrote a short report with George Gamow entitled ``Theory of the Detonation Process".
\newline

At about that time Einstein had agreed to serve as a consultant to our group but did not want to travel to Washington. So there had to be a liaison person and I was given that opportunity. Since Einstein did not know me, there had to be someone to introduce us. It then happened that I was introduced to Einstein by John von Neumann, one of the most important mathematicians of all time, and who had also become a consultant to our group. It was a very great experience for a new Ph.D. to be introduced to Einstein by Von Neumann! During the following period I met Einstein every week until Gamow joined our group and became the liaison person.
\newline

When peace finally came, Bob Sachs, whom I had come to know during the war, invited me to join the nuclear theory group of which he had become the head, at what became the Fermi Lab at the University of Chicago. I was there long enough to enjoy meeting some members of the famous postwar class, particularly Jack Steinberger and Frank (C.N.) Yang and of course Fermi, to learn some nuclear physics, and to publish a paper in the Physical Review; but my real interest was not nuclear physics, but more fundamental physics. So I optimistically wrote to Oppenheimer and to Pauli to apply for a postdoc, since they had begun to work on meson theory. When Oppenheimer accepted me, the Yukawa theory was still very young, and mesons coming from an accelerator had yet to be observed. The project that he proposed to me led to the paper "The Gamma Instabilities of Mesons". This work was carried out with the methods available at the time, before Schwinger, Tomonaga, and Feynman had made peace with the infinities. (This calculation was partially repeated by Jack Steinberger sometime later, with Feynman diagrams, but with the same physical picture.)
\newline

About a year after I "blew in" as he described it, Oppenheimer accepted the directorship of the Institute for Advanced Studies in Princeton. When he moved to Princeton, he brought his Berkeley group with him (H. Lewis, S. Wouthuysen, L. Foldy, and myself). My friends at the Institute that year were the Berkeley group and the Dutch friends of Sieg Wouthuysen including Abraham Pais, the mathematician Nicolaas Kuiper, and his wife Agnete, the daughter of H.A. Kramers.
\newline

During this period at the IAS, I began to study the quantization of unitary field theories and was encouraged by the interest of C. M\o ller in Copenhagen. I also had some productive interaction with David Bohm before he was fired from his position as a young Assistant Professor as part of the McCarthy witch hunt. The following year Oppenheimer obtained a position for me at Caltech as a Junior Fellow working with Bob Christy. That same year Christy and I welcomed Yukawa at LAX on his first postwar visit to this country.
\newline

At Caltech I gave the first postwar course in field theory based on what I had learned from the written presentations of Pauli and Wentzel. There I met and began to work with Mal Ruderman, and he became my first doctoral student. That year UCLA offered positions to Gamow and myself. Gamow was happy about going to UCLA when he visited me in my office at Caltech but for reasons that I never understood, Hans Bethe influenced UCLA to withdraw the offer to Gamow. That was a big disappointment for me but I was persuaded by David Saxon that there was a bright future for UCLA, and so Saxon and I became what was then the high energy theory group.
\newline

During this period I attended two sessions of the Michigan summer school at which Schwinger (1948) and Feynman (1949) described their respective reformulations of QED. Schwinger's was deeper and more complete while Feynman's was easier to use but at that time incomplete. One may give a feeling for the impact of Schwinger by quoting Dyson who wrote home that "in a few months we shall have forgotten what pre-Schwinger physics was like." Bethe at that time described this period as the most exciting in physics since the great days of 1925-30 when quantum mechanics was being discovered.
\newline

By this time the ``Universal Fermi Interaction" (UFI) had been proposed by Wheeler and Tiomno. Adopting the UFI, Mal Ruderman and I calculated (1949) $\pi e / \pi \mu$ in the same way, i.e. pre-Schwinger, that I had previously calculated the $\pi \gamma$ rates. The main uncertainties in these calculations were the tensor natures of the pi meson and the beta interaction. To agree with experiment, our results showed that the pi had to be a pseudoscalar and the beta interaction had to be a partly axial vector. The following year Panofsky established the pion as a pseudoscalar. The other result that the beta interaction had to be partial axial vector was confirmed after the discovery of parity violation.
\newline

After the encouraging tests of the UFI, I proposed a unitary field theory describing fermions obeying the Dirac equation and interacting via the UFI. Then Mal and I, with Bob Lelevier, discovered that this model had particle-like eigensolutions of a nonlinear differential equation. In other contexts, particularly Yang Mills, these were later called solitons. Feynman's comment to me at that time was, ``I see how you got into this, but I don't know how you will get out of it." The same Lagrangian was proposed by Heisenberg and Pauli a little after us. I continued to study similar models of non-linear differential equations with Steve Gasiorowicz (1954) and Peter Kaus (1954) and Chris Fronsdal (1956). (Shortly afterwards Chris began his pioneering use of Lie algebras in particle physics.) 
\newline

In 1956 Norma and I were married. We had previously met on my first visit to Princeton where she was then employed by the meteorologist, Jule Charney, and by John von Neuman as a member of the group that built the johniac (which is now in the Smithsonian). Norma is a violinist who has led a second life in chamber music with players from a global directory that included some physicists found at the IAS in Princeton and at CERN in Geneva.
\newline

At a slightly earlier time I had made an exploratory calculation of mesonic corrections to beta decay to account for the difference between the Fermi and Gamow-Teller coupling constants that had been pointed out to me by Steve Moskowski (1954). This was followed by a similar but more detailed calculation with Ralph Behrends of the radiative corrections to muon decay (1955). At this time Alberto Sirlin ``blew in" from Argentina, and shortly thereafter appeared the paper ``Radiative Corrections to Decay Processes" written by the three of us (1956).
\newline

Alberto, who had been learning quantum mechanics from Feynman in Brazil, arrived at about the same time as Chris arrived from the Bohr Institute in Copenhagen. My own travels began immediately after I came to UCLA. On my first semi-sabbatical I was invited to Copenhagen by C M\o ller, where I met Harry Lehmann. On every following semi-sabbatical I tried to go to either the Institute in Princeton or to Cern in Geneva. At Princeton I made many friends, especially including Yang, whom I had known in Chicago, Nambu, and Zumino. Although I managed to bring Lehmann, Symanzik, Yang, and Nambu to UCLA for visits, none of these visitors was finally persuaded to stay. With Ferrara we were partially successful!
\newline

Meanwhile the ``zoo of elementary particles" and the symmetries of their associated fields was growing so that Einsten's dream of understanding them in the context of a unified field theory was becoming more remote. Einstein himself attempted to generalize the spacetime connection to include the Maxwell field. I tried to follow the spirit of this approach by defining a new spacetime connection lying in the Lie algebras of the newly discovered particles. This approach necessarily leads to theories of the Einstein or Yang-Mills variety. The Yang-Mills paper was published in 1954. I published my own speculations in ``Spacetime of the Elementary Particles" in 1960 and in ``Elementary Interactions in Spaces of Constant Torsion" in 1961 followed by a more systematic study of ``The Weak and Strong Interactions and General Covariance" with Bill Ramsay in 1963.
\newline

At that time, however, my main interests were confined to the weak interactions and the possibility of understanding the UFI. Like many others I felt that the UFI was mediated by a massive vector and I began to construct effective gauge theories of massive vectors, first alone, next with Larry Staunton, and then with Jan Hilgevoord, a visitor from Amsterdam, with the titles ``Non-linear Pion Nucleon Lagrangian" (1969), ``Massive Gauge Fields and Associated Pseudoscalar Mesons" (1969), ``Effective Lagrangians Associated with Massive Gauge Fields" (1970).
\newline

These effective field papers and others of that period described non-abelian or Yang-Mills theories that were in encouraging but rough agreement with strong coupling experiments and the prevailing current algebra, but ours were technically different in being implemented not on Lie algebras but on the corresponding group space (a formalism that I had previously used to extend the work of V. Bargmann on the quantum symmetries of the hydrogen atom and that I had used with Don Levy in a reformulation of scattering theory). Prominent among the effective field theories were the papers of Schwinger and of Weinberg. In his Nobel Prize lecture Steve Weinberg noted that his strong coupling Yang-Mills paper had the essence of the electroweak theory, but was applied to the wrong particles. 
\newline

Then in 1971 Dave Saxon finally succeeded in persuading Julian Schwinger to leave Harvard and come to UCLA. He came with three post-docs: Kim Milton, Lester de Raad, and Wu-Yang Tsai. They were then beginning to make applications of Source Theory, a formalism that Julian had recently constructed as an infinity-free replacement of the monumental operator field theory he had previously created. I joined the Source Theory lunch group and continued to lunch with Julian after the postdocs had left. I also proposed to my students, Joel Kvitky and Jan Smit, that they examine massive vector theories as Source Theories. Jan did a beautiful job, but was not too enthused about it because at the time he was writing his thesis, he was also working out the basics of lattice field theory (earlier and independently of Alexander Polyakov and Kenneth Wilson). I hope that the Source Theory of Julian Schwinger has the same promising, if delayed, future as the Green's Functions of George Green has had, thanks to Julian.
\newline

In 1976 Supergravity was discovered by Ferrara, Freedman, and van Nieuwenhuizen, and by Deser and Zumino. Schwinger had in 1966 laid the basis for supersymmetry in the first volume of his trilogy ``Particles, Sources, and Fields" but did not pursue supersymmetry further. In a short paper (1979) I showed how one could reach supergravity from Schwinger's formulation of supersymmetry via Fermi-Bose transformations, and in a longer paper with Milton and Urrutia, it was shown how local supersymmetry alone was sufficient to yield full supergravity.
\newline

Then in 1985 we were fortunate in being able to persuade Sergio Ferrara to move from CERN to UCLA for one quarter per year. Although Sergio was here only for one quarter, I found his yearly visits enormously stimulating, especially since he brought with him the group with whom he had been working that year. I was greatly interested in the developments in supergravity during the nineteen eighties. Then I was mandatorily retired in 1986 when I became 70. (The law was changed that year and now there is no mandatory retirement.) I continued to work on supergravity and related topics with Manuel Villasante and Jaegu Kim.
\newline

I also began a new program in an effort to construct a field theory based on a local affine (Kac-Moody) algebra. This work was carried out with Cristina Cadavid. Since our aim was to find the space-time connection ``that nature prefers", the general goal was to prepare a contest of models from which experiment would choose. Continuing along these lines we next began to move further from Lie algebras to SLq(2), the knot algebra, beginning in the early 1990s. At the same time after the theoretical discovery of electroweak theory and the experimental discovery of the Z particle, there was general acceptance of the ``Standard Model". Around 2004 I realized that there was a very interesting extension of the standard model if the elementary particles were knotted.
\section*{Some Properties of the Knot Model[1]}
There is a Lagrangian dynamics underlying the knot extension, and the physical picture is very intuitive as follows: the elementary particles are knotted current loops of energy-momentum in the sense of SLq(2), i.e. with the topological quantum numbers $(j,m,m')$ determined by
\begin{align*}
(j,m,m') = \frac{1}{2}(N,w,r+o)
\end{align*}
where $(N,w,r) = (\text{number of crossings}, \text{writhe}, \text{rotation})$ respectively that describe the 2d-projection of the corresponding classical knot, and where $o$ is an odd integer set equal to unity for the simplest knot, the trefoil. This equation restricts the kinematics of the quantum dynamics by relating every state of the quantum knot to a corresponding classical knot.
\newline

There is also an empirical relation
\begin{align*}
(j,m,m') = 3(t,-t_3, -t_0)_L
\end{align*}
connecting the quantum trefoils $(j,m,m')$, describing the leptons and quarks, with the left-chiral fields $(t, -t_3, -t_0)_L$ of these particles.
\newline

The fermion and boson knots are distinguished by having odd and even numbers of crossings respectively. The leptons and quarks are the simplest quantum knots, the quantum trefoils with three crossings and $j=\frac{3}{2}$. At each crossing there is a preon. The free preons are twisted loops with one crossing and $j=\frac{1}{2}$. The $j=0$ states are simple loops with zero crossings.
\newline

The preon content of the quarks matches unexpectedly and exactly with the model of Harari (1979) and of Shupe (1979) and of course agrees with the experimental basis of their models.
\newline

The $j=0$ state suggests a cosmological model of the early universe beginning at $t=0$ with a gas of $j=0$ particles which may be called ``yons" (after ylem, an accepted term for primordial matter). In this gas there would be some gravitationally induced inelastic collisions between yons, simple loops with no crossings, resulting in preons, twisted loops with one crossing. A fraction of the preons $(j=\frac{1}{2})$ would in turn combine to form two preon $(j=1)$ and three preon $(j=\frac{3}{2})$ states with two and three crossings respectively. The $j = \frac{3}{2}$ states would be recognized in the present universe as leptons and quarks.
\newline

Since the $j = \frac{1}{2}$ and $j=1$  particles with one or two crossings, respectively, are not topologically stable in three dimensions and can relapse into a $j=0$ state with no crossings, the building up process does not produce topologically stable particles before the $j = \frac{3}{2}$ leptons and quarks with three crossings are reached.
\newline

If this evolutionary process is now incomplete, the residual yon gas could outweigh visible matter, and could be called ``dark energy" or as Shakespeare asked, ``What light through yon(der) window breaks"\cite{ruth}.

\pagebreak

%abers, patrice, cristina


\begin{thebibliography}{9}
\bibitem{slq2}
Phys. Rev. D89 (2014),
Int. J. Mod. Phys. A30 (2015),
arXiv:1511.07919
\bibitem{ruth}
Ruth Finkelstein, private communication
\end{thebibliography}
\end{document}